\documentclass[aps,preprint,showpacs,superscriptaddress]{revtex4}  

\usepackage{graphicx}  
\usepackage{dcolumn}   
\usepackage{bm}        
\usepackage{amssymb}   

\newcommand{\feo}{Fe$_3$O$_4$}

\begin{document}

\title{Field-dependent anisotropic magnetoresistance and planar Hall effect in epitaxial magnetite thin films}

\author{N. Naftalis}
\email{naftaln@mail.biu.ac.il}
\author{A. Kaplan}
\author{M. Schultz}

\affiliation{Department of Physics, Nano-magnetism Research Center, Institute of Nanotechnology and Advanced Materials, Bar-Ilan University, Ramat-Gan 52900, Israel}

\author{C.~A.~F.~Vaz}
\author{J. A. Moyer}
\author{C. H. Ahn}

\affiliation{Department of Applied Physics, Yale University, New
Haven, Connecticut 06520, USA}

\author{L. Klein}

\affiliation{Department of Physics, Nano-magnetism Research Center, Institute of Nanotechnology and Advanced Materials, Bar-Ilan University, Ramat-Gan 52900, Israel}

\pacs{75.47.-m, 75.50.Bb, 75.70.Ak, 72.15.Gd}

\begin{abstract}

A systematic study of the temperature and magnetic field dependence of the longitudinal and transverse resistivities of epitaxial thin films of magnetite (\feo) is reported. The anisotropic magnetoresistance (AMR) and the planar Hall effect (PHE) are sensitive to the in-plane orientation of current and magnetization with respect to crystal axes in a way consistent with the cubic symmetry of the system. We also show that the AMR exhibit sign reversal as a function of temperature, and that it shows significant field dependence without saturation up to 9 T. Our results provide a unified description of the anisotropic magnetoresistance effects in epitaxial magnetite films and illustrate the need for a full determination of the resistivity tensor in crystalline systems.
\end{abstract}

\maketitle

\section{Introduction}

Magnetite (\feo) is a 3d transition-metal oxide and an itinerant ferrimagnet with a high Curie temperature (858 K) and intriguing transport properties, including a metal insulator transition at T $\sim$ 120 K (the Verwey transition) and high spin polarized current attributed to its being half metallic. The latter property has given rise to a renewed interest in this material as a promising material system for spintronics applications \cite{magnetite_overview,magnetite_vareway} and special efforts have been devoted to elucidate its magnetotransport properties.

Two of the most fundamental magnetotransport properties of itinerant ferromagnets are the dependence of the longitudinal ($\rho_\mathrm{long}$) and transverse resistivity ($\rho_\mathrm{trans}$) on the relative orientation of the magnetization (\textbf{M}) and electric current (\textbf{J}), known as anisotropic magnetoresistance (AMR) and planar Hall effect (PHE), respectively. Both phenomena are attributed to spin-orbit interaction which mixes spin-up and spin-down states \cite{AMR_Smit}. Phenomenologically, AMR and PHE are commonly expressed as \cite{old_AMR}:
\begin{eqnarray}
\label{Eq:rhoxx_old}
\rho_\mathrm{long} & = & \rho_{\perp} + (\rho_{\parallel} - \rho_{\perp})\cos^{2} \varphi\\
\label{Eq:rhoxy_old}
\rho_\mathrm{trans} & = & (\rho_{\parallel} - \rho_{\perp}) \sin \varphi \cos \varphi,
\end{eqnarray}
where $\varphi$ is the angle between \textbf{J} and \textbf{M}, and $\rho_{\parallel}$ and $\rho_{\perp}$ are the longitudinal and transverse resistivities corresponding to \textbf{M} $\parallel$ \textbf{J} and \textbf{M} $\perp$ \textbf{J}, respectively. While Eqs. \ref{Eq:rhoxx_old} and \ref{Eq:rhoxy_old} are applicable for isotropic or polycrystalline materials, they fail to fully account for the magnetotransport properties of crystalline systems, where additional terms arise \cite{doring}. This has been demonstrated for [001]-oriented epitaxial manganites thin films \cite{LSMO_Yosi,LCMO_Nati}, where a description of AMR and PHE is provided by taking into account the cubic symmetry of the system. The applicable equations are:
\begin{eqnarray}
\label{Eq:rho_par1}
\rho_\mathrm{long} & = & A\cos(2\alpha-2\theta)+B\cos(2\alpha+2\theta)+C\cos(4\alpha)+D\\
\label{Eq:rho_par2}
\rho_\mathrm{trans} & = & A\sin(2\alpha-2\theta)-B\sin(2\alpha+2\theta),
\end{eqnarray}
where $\alpha$ and $\theta$ are the angles between \textbf{M} and \textbf{J} relative to the [100] crystal direction, respectively (see inset of Fig. \ref{Fig:RvsT}).

In this work we focus our attention on the anisotropic magnetoresistance behaviour of epitaxial magnetite films, where the need for an additional term with fourfold symmetry has been suggested in the literature \cite{fe3o4_ramos_prb,fe3o4_apl,fe3o4_jap}. We show, however, that Eqs. \ref{Eq:rho_par1} and \ref{Eq:rho_par2} are required to fully describe the AMR and PHE of cubic epitaxial films for arbitrary directions of current and magnetization.

We find different temperature and magnetic field dependence of the parameters $A$ and $B$, suggesting that they originate from different scattering mechanisms. The parameter $D$ is found to exhibit a strong magnetic field dependence, which is ascribed mainly to the effect of antiphase boundaries in the \feo\ film. The full description of the AMR and PHE behavior of magnetite films allows new insight on the transport mechanisms in this compound and enables the optimization of the planar Hall voltage response, which could be useful for non-volatile memory devices based on the planar Hall effect \cite{BKY04,BKY06,BKW07}.

\section{Experiment}

For this study, \feo\ films $\sim 20$ nm thick were grown on MgO(001) single crystals by molecular beam epitaxy in an ultrahigh vacuum deposition system with a base pressure of $1\times 10^{-9}$ mbar. Prior to film growth the MgO substrate was cleaned under an oxygen plasma at 520 K for 30-60 min, which renders the surface well ordered and free of impurities, as determined by surface electron diffraction (low energy electron diffraction, LEED, and reflection high energy electron diffraction, RHEED) and x-ray spectroscopy (Auger electron spectroscopy, AES, and x-xay photoemission spectroscopy, XPS), respectively. For the magnetite growth, an atomic Fe beam was thermally generated from an effusion cell under an O$_2$ partial pressure of $2\times 10^{-7}$ mbar, with the substrate temperature held at 520 K. Film thickness was estimated from a calibrated thickness monitor, while film crystallinity was monitored during growth using RHEED. After growth, the film was characterized {\it in situ} by RHEED and XPS, showing that the magnetite films are single crystalline and stoichiometric \cite{Anisotropy_field1}. The MgO substrate crystal orientation was confirmed by Laue diffraction. The films were subsequently patterned by photolithography for longitudinal and transverse resistivity measurements along different angles ($\theta$) of the current with respect to the magnetite crystal axes.

\section{Results and Discussion}

\begin{figure}
\includegraphics[scale=0.5]{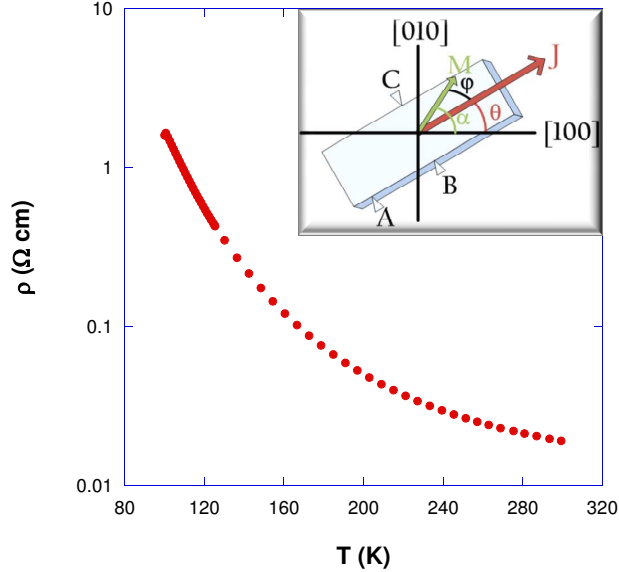}
\caption{Resistivity ($\rho$), in log scale, as a function of temperature. Inset: Sketch of the relative orientations of the current density \textbf{J}, magnetization \textbf{M}, and the crystallographic axes. $\rho_\mathrm{long}$ is measured between A and B, $\rho_\mathrm{trans}$ is measured between B and C.}
\label{Fig:RvsT}
\end{figure}

The temperature dependence of the resistivity is presented in Fig.~\ref{Fig:RvsT} in semi-log scale. We note a change in the slope of the resistivity when approaching the Verwey transition at T$\sim 120$ K, indicative of the high quality of the films. Thin magnetite films tend to have a less pronounced resistivity discontinuity at the Verwey transition as compared with bulk magnetite consistent with earlier transport studies of \feo\ \cite{ZB00,Verwey_film}. The data in this graph essentially corresponds to the temperature dependence of the $D$ term in Eq. \ref{Eq:rho_par1}.

\begin{figure}
\includegraphics[scale=0.7]{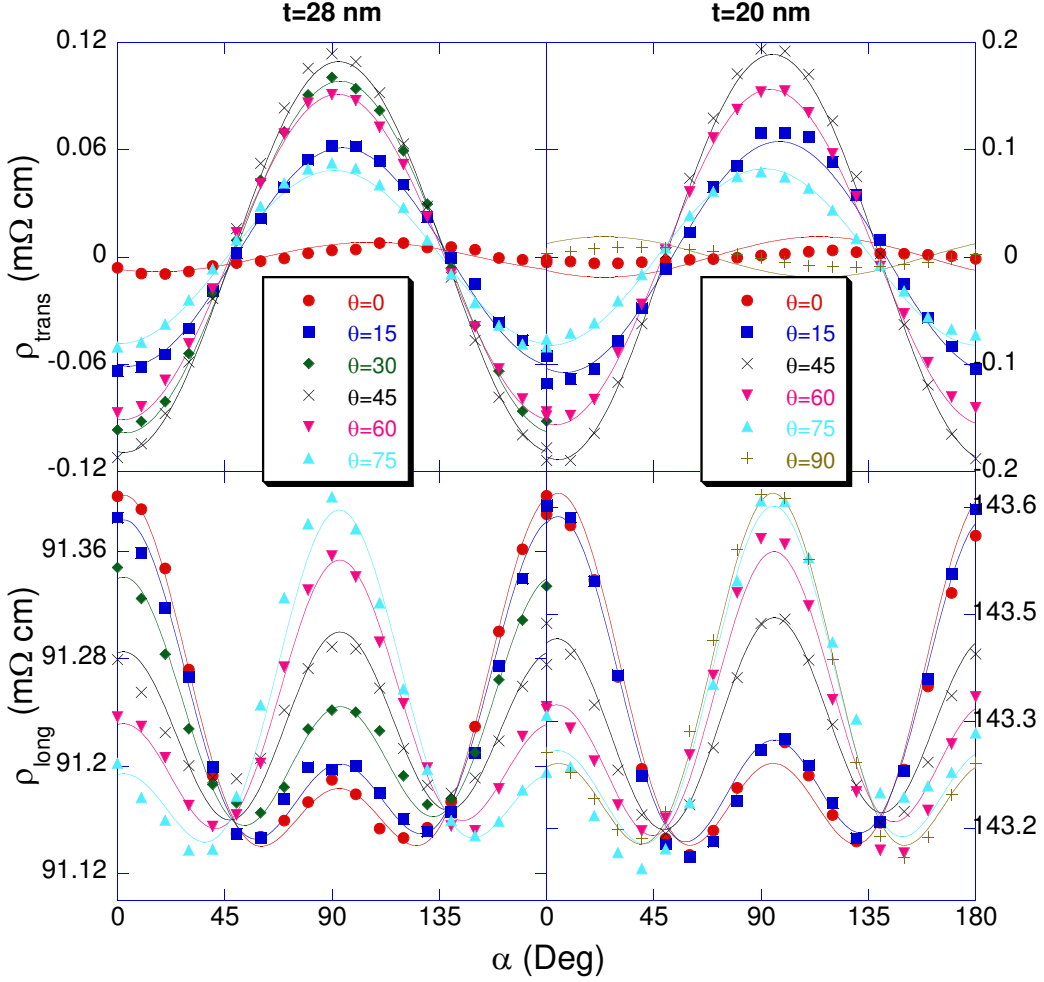}
\caption{$\rho_\mathrm{long}$ and $\rho_\mathrm{trans}$ vs $\alpha$, the angle between the magnetization \textbf{M} and [100], for different angles $\theta$ (the angle between the current direction \textbf{J} and [100]) at T=150 K with an applied magnetic field of 9 T for two different samples. The solid lines are fits to Eqs. \ref{Eq:rho_par1} and \ref{Eq:rho_par2}.}
\label{Fig:AMR&PHE}
\end{figure}

The variation of $\rho_\mathrm{long}$ and $\rho_\mathrm{trans}$ as a function of the angular direction of the applied magnetic field ($\alpha$, relative to [100]\feo) was measured for six different device structures with varying angle $\theta$ between the current density direction and the [100] direction of the \feo\ film. The measurements were performed with a constant magnetic field of up to 9 T rotating in the plane of the film. This field is much larger than the reported in-plane magnetization saturating field \cite{Anisotropy_field1,Anisotropy_field2}; hence, the magnetization is expected to be practically parallel to the applied field. Two representative examples of the angular variation of $\rho_\mathrm{long}$ and $\rho_\mathrm{trans}$ are shown in Fig.~\ref{Fig:AMR&PHE} for $T=150$ K. We find that while $\rho_\mathrm{trans}$ varies sinusoidally, its amplitude depends on $\theta$ and the location of the extremal points depend on $\alpha$, inconsistent with Eq. \ref{Eq:rhoxy_old}, according to which the amplitude is $\theta$-independent. Furthermore, $\rho_\mathrm{long}$ lacks twofold symmetry. These observations clearly show that Eqs. \ref{Eq:rhoxx_old} and \ref{Eq:rhoxy_old} fail to describe AMR and PHE in epitaxial films of magnetite. On the other hand, Eqs. \ref{Eq:rho_par1} and \ref{Eq:rho_par2}, which take into account the crystal symmetry of magnetite, provide a good description for the angular dependence of $\rho_\mathrm{long}$ and $\rho_\mathrm{trans}$, as shown by the fits to the experimental data. The equations have three parameters and in the fitting process we use the same values for fitting the AMR and PHE curves for the six current directions (see Fig.~\ref{Fig:AMR&PHE}). We note that Eqs. \ref{Eq:rho_par1} and \ref{Eq:rho_par2} provide a good fit to the data even below the Verwey temperature, where bulk magnetite is monoclinic. This is expected since our films are epitaxial and they maintain cubic symmetry in the plane of the magnetization and current.

\begin{figure}
\includegraphics[scale=0.5]{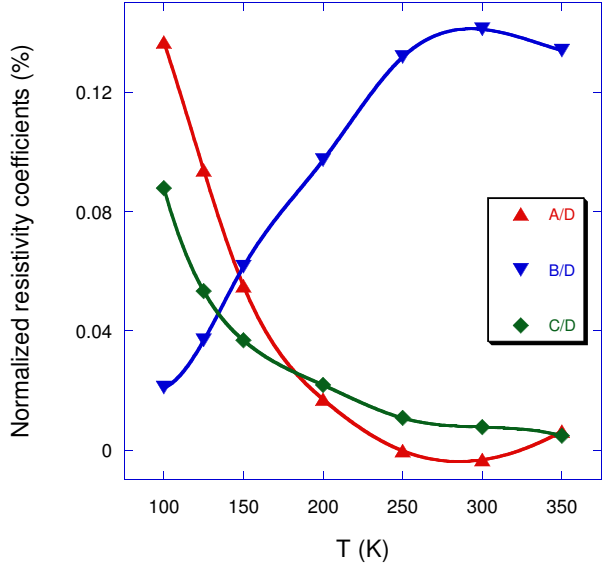}
\caption{The parameters in Eqs. \ref{Eq:rho_par1} and \ref{Eq:rho_par2} normalized by the parameter $D$ as a function of temperature with an external field of 4 T. The lines are guides to the eye.}
\label{Fig:MRvsT}
\end{figure}

The temperature dependence of the three fitting parameters, $A$, $B$, and $C$, is shown in Fig.~\ref{Fig:MRvsT}. We note that at high temperatures $B$ is dominant, while at low temperatures $A$ and $C$ dominate. The fact that the coefficients $A$ and $B$ show distinct temperature dependencies indicate that they have different origins. The terms with the parameter $A$, as their coefficient are identical to the terms that appear in Eqs. \ref{Eq:rhoxx_old} and \ref{Eq:rhoxy_old}, and they are independent of crystal symmetry. On the other hand, the terms with the parameters $B$ and $C$ as their coefficients are sensitive to the crystal symmetry. The parameter $C$ appears only in the equation for $\rho_\mathrm{long}$ and is responsible for a fourfold angular symmetry which provides a clear manifestation of crystal symmetry effects. The parameter $B$ contributes to both $\rho_\mathrm{long}$ and $\rho_\mathrm{trans}$ with a twofold symmetry, and the crystal symmetry is manifested when AMR and PHE are measured for current directions away from the principal axes. This may explain why the crystal contribution to AMR and PHE at high temperatures has not been reported before \cite{fe3o4_ramos_prb}. We note also that, according to Eqs. 3 and 4, the anisotropic magnetoresistance as usually defined, $\mathrm{AMR} = \rho(\bf{M}\parallel J) - \rho(\bf{M}\perp \bf{J})$, yields $\mathrm{AMR} = 2A + 2B\cos4\theta$, showing that for $B \geq A$ the AMR can be positive or negative, depending of the direction of the electric current ($\theta$). In particular, for $\theta = \pi/4$, $\mathrm{AMR} = 2(A-B)$, and the AMR, as conventionally defined, changes sign at around 150 K.

\begin{figure}
\includegraphics[scale=0.5]{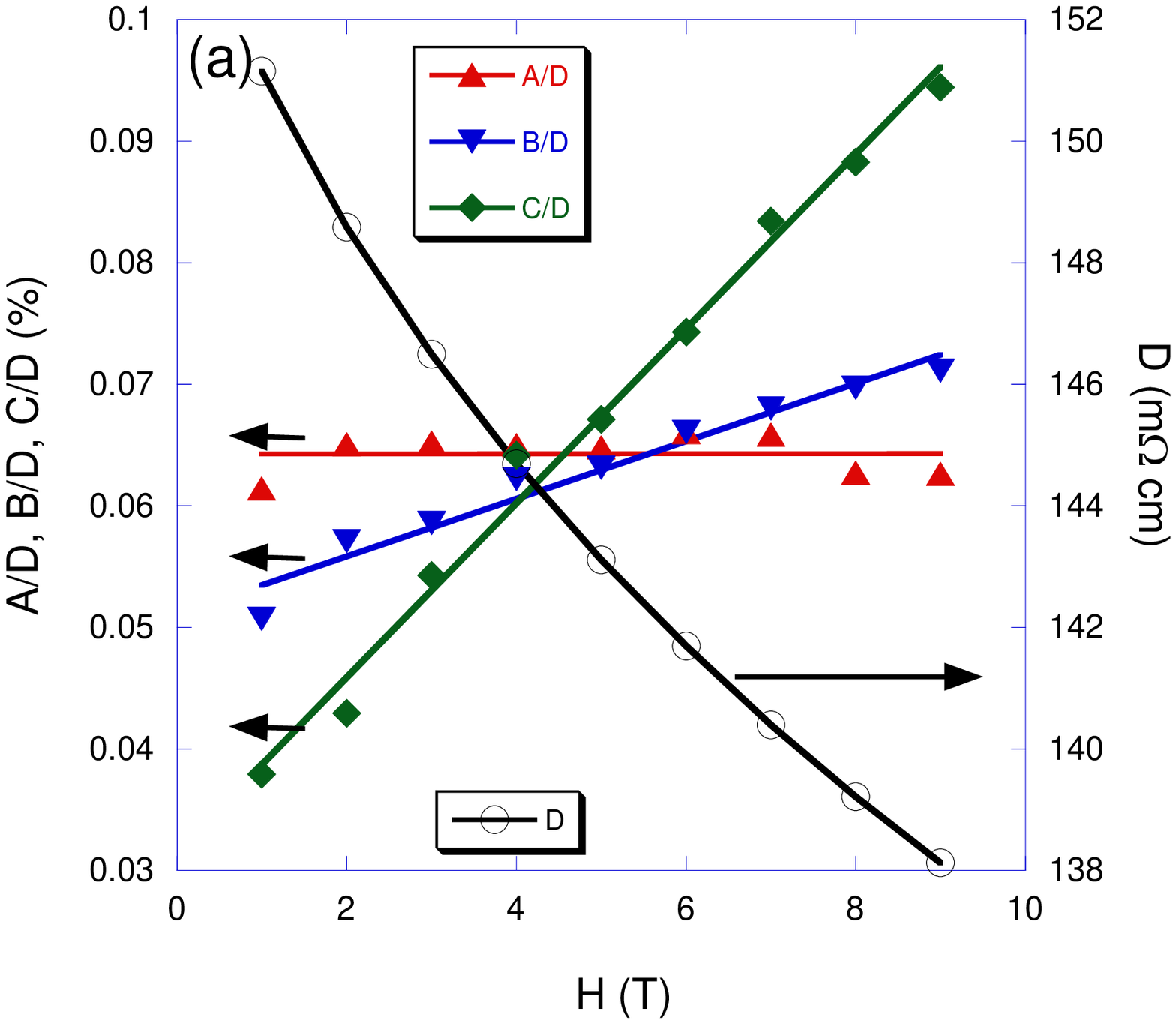}
\includegraphics[scale=0.5]{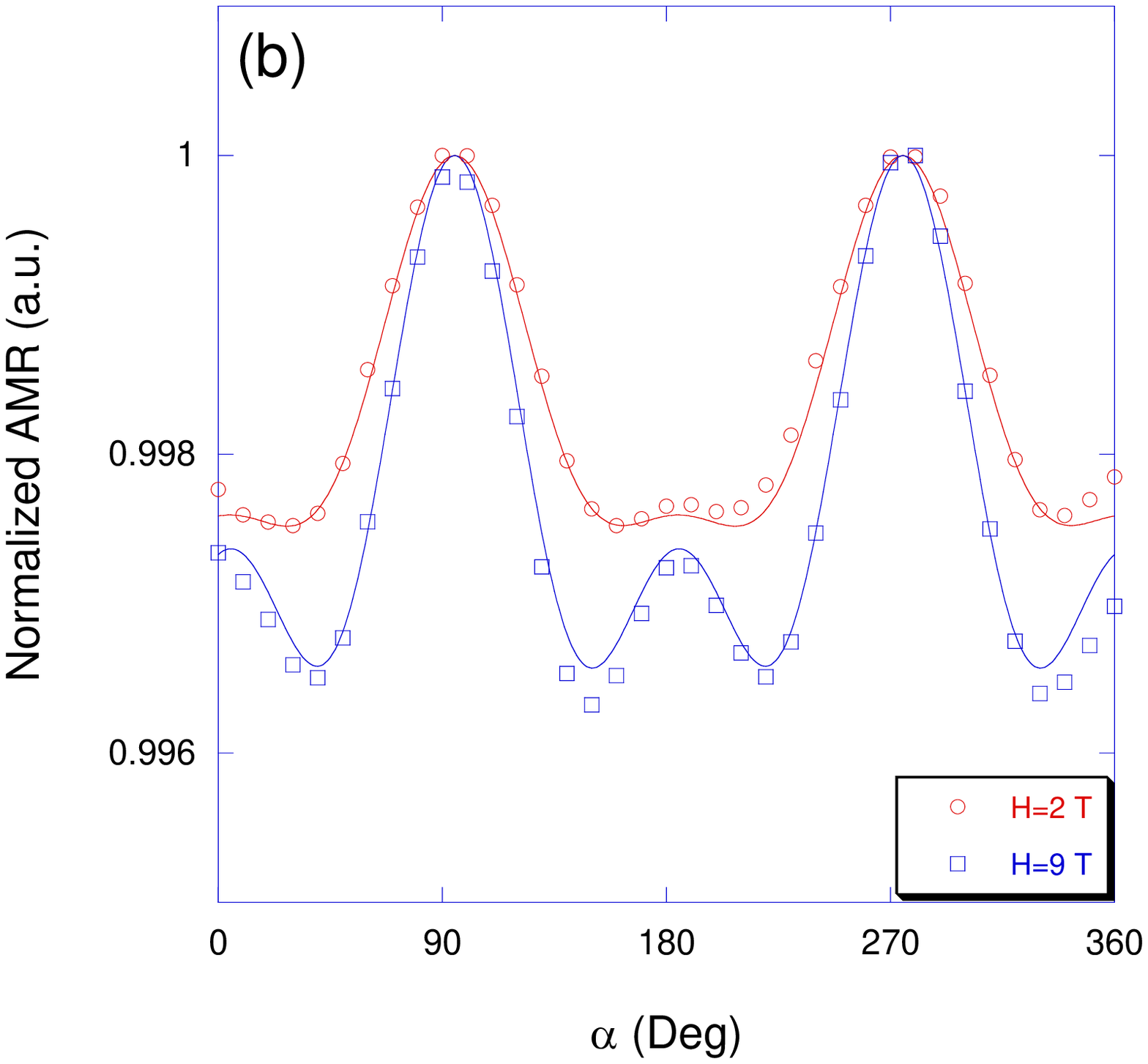}
\caption{(a): (right axis) The parameter $D$ and (left axis) the other parameters from Eqs. \ref{Eq:rho_par1} and \ref{Eq:rho_par2} normalized by $D$ as a function of the external field at T=150 K. The lines are guides to the eye. (b): AMR normalized to its maximum value as a function of $\alpha$ for two different external magnetic fields at $T=150$ K. The solid lines are fits to Eq. \ref{Eq:rho_par1}.}
\label{Fig:MRvsH}
\end{figure}

The field dependence of the magnetoresistance fitting parameters was also measured at selected temperatures and representative results for one sample are shown in Fig.~\ref{Fig:MRvsH}(a). While the field dependence of $A$ and $B$ is weak, the parameter $C$ varies strongly with the magnetic field amplitude, with no sign of saturating up to 9 T. The field dependence yields quite dramatic changes in the AMR as can be seen in Fig.~\ref{Fig:MRvsH}(b), which exhibits a much stronger fourfold symmetry component when the field is increased from 2 T to 9 T. The field dependence of the magnetoresistance parameters varies with temperature, but the qualitative behavior is similar to that shown in Fig.~\ref{Fig:MRvsH}(a).

The parameter $D$, which represents magnetoresistivity term insensitive to $\alpha$ and $\theta$, is also found to vary strongly with the magnetic field, giving rise to large magnetoresistances of the order of --10\% at 9 T. While the magnetoresistance ratio of bulk magnetite is relatively weak, of the order of --0.1\% \cite{ZB00}, much larger effects, of up to --10\% at 5 T \cite{calcAPB}, are observed in thin films.

\begin{figure}
\begin{center}
\includegraphics[scale=0.5]{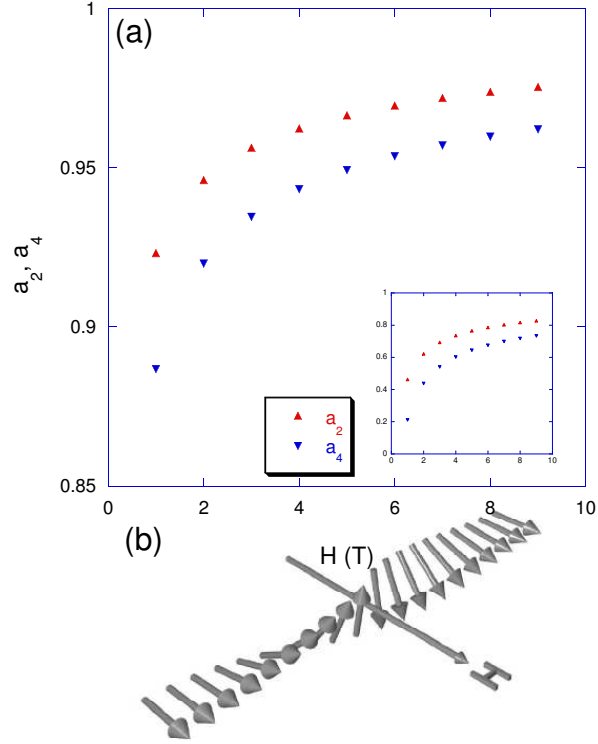}
\end{center}
\caption{The field dependence of $a_2$ (the numerical average of $\cos 2 \beta$) and $a_4$ (the numerical average of $\cos 4 \beta$) for the whole domain and (Inset) for the vicinity of the boundary alone (see text). (b): an illustration of spin orientation of a ferromagnetic chain with anti-ferromagnetic coupling at the origin, taken from \cite{calcAPB}.}
\label{Fig:CvsH}
\end{figure}

Several different factors may affect the magnetotransport behavior of magnetite, including the spin-orbit coupling, electron localization on approaching the Verwey transition, changes in the spin polarization as a function of temperature \cite{VAH09}, and changes in the character of the charge carrier at temperatures near room temperature \cite{new}. The combination of all these factors makes a correct identification of the dominant scattering processes in play challenging. We do note, however, that the difference between bulk compounds and thin films grown on MgO (particulary, lack of high field saturation) has been attributed to the presence of antiphase boundaries (APB) \cite{APB,ZB00,calcAPB,Sofin_MR_APB,APDdistribution}. The APB disturb the magnetic configuration even at high magnetic fields, particularly close to the boundary, as illustrated in the inset of Fig. \ref{Fig:CvsH}(b); hence, they are likely to be the origin of the magnetic field dependence of the parameter $D$. We turn now to examine whether there could be a link between APB and the observed field dependance of AMR.

We follow previous treatments and calculate the spin configuration, $\beta (x)$, in an infinite ferromagnetically-coupled spin chain with anti-ferromagnetic boundary at the origin. The energy per unit area of such a spin chain to the left to the origin is given by \cite{1Denergy}
\begin{equation}
\varepsilon=\int_{-\infty}^0[-M_sH\cos\beta+A_F(\frac{d\beta}{dx})^2]dx
\label{Eq:spin_energy1}
\end{equation}
where the first term in the integral is the Zeeman energy and the second term is the nearest neighbor exchange contribution. The angle between the magnetization, $\it{M_s}$, and the magnetic field, $\it{H}$, is denoted by $\beta$, and $\it{A_F}$ is the exchange stiffness constant. We neglect the magnetocrystalline, since it is much smaller than the external field that we used in our measurements \cite{Anisotropy_field1,Anisotropy_field2}. Standard variational calculus procedures yield:
\begin{equation}
\beta(x)=4\arctan(C e^{x/x_0})
\label{Eq:spin_arrange}
\end{equation}
where $x_0^2=2A_F / M_sH$, $C=\tan(\beta_{AF}/4)$ and $2\beta_{AF}$ is the angle between the spins on both sides of the boundary which depends on the anti-ferromagnetic coupling, and can be calculated based on the equilibrium condition of the chain \cite{calcAPB}. We note that this model does not take into account the distribution in the orientation of the APB, since $\beta (x)$ does not depend on the angle between the APB and magnetic field.

Figure \ref{Fig:CvsH} presents the spatial average of $\cos 2 \beta$ and $\cos 4 \beta$ denoted as $\langle\cos 2 \beta\rangle$ and $\langle\cos 4 \beta\rangle$, respectively, using a ferromagnetic coupling of 3 K and anti-ferromagnetic coupling of -22 K \cite{exchange}. The averages are $a_2 = \frac{1}{2n+1}\sum_{j=-n}^{j=n} \cos (2\beta(jd)) $ and $a_4 = \frac{1}{2n+1}\sum_{j=-n}^{j=n} \cos (4\beta(jd))$ respectively, where $d$ is the separation between spins (about 1 {\AA} \cite{APB}) and the domain size is $\sim$ 350 {\AA} \cite{domain_size}. Changing the domain size in a reasonable range between 250 {\AA} and 350 {\AA} does not affect the results significantly.

The field dependence of $a_2$ and $a_4$ is much smaller than the field dependance of $B$ and $C$. However, as suggested before \cite{calcAPB}, the largest contribution to resistivity is close to the boundary (due to the significant change in the spin orientation) and taking the average only in the vicinity of the boundary yields a more dramatic field dependance (see inset of Fig. \ref{Fig:CvsH}). Thus, one can not exclude that a more sophisticated model that would consider the effect of spin variation on $\rho$ would yield better agreement with experiment.

In summary, we measured the temperature and magnetic field dependence of the longitudinal and transverse resistivities of \feo\ as a function of the current direction relative to the crystal axes.
The results shed light on the interplay between magnetism and electrical
transport in this class of materials and may serve as a basis
for further study of the microscopic origin of magnetotransport properties of magnetite.

\begin{acknowledgements}
L.K. acknowledge support by the Israel Science Foundation founded by the Israel Academy of Science and Humanities (Grant 577/07). Work at Yale supported by NSF under Grants No.
DMR 1006256 and No. DMR 0520495, and FENA.
\end{acknowledgements}


\begin{thebibliography}{99}
\bibitem{magnetite_overview} M. Ziese, Rep. Prog. Phys. \textbf{65}, 143 (2002).
\bibitem{magnetite_vareway} F. Walz, J. Phys.: Condens. Matter. \textbf{14}, R285 (2002).
\bibitem{AMR_Smit} J. Smit, Physica (Amsterdam) \textbf{17}, 612 (1951).
\bibitem{old_AMR} T. R. McGuire and R. I. Potter, IEEE Trans. Magn. \textbf{11}, 1018 (1975).
\bibitem{doring} W. D\"{o}ring, Ann. Phys \textbf{424}, 259 (1938).
\bibitem{LSMO_Yosi} Y. Bason, J. Hoffman, C. H. Ahn, and L. Klein, Phys. Rev. B \textbf{79}, 092406 (2009).
\bibitem{LCMO_Nati} N. Naftalis, Y. Bason, J. Hoffman, X. Hong, C. H. Ahn, and L. Klein, J. Appl. Phys \textbf{106}, 023916 (2009).
\bibitem{fe3o4_ramos_prb} R. Ramos, S. K. Arora, and I. V. Shvets, Phys. Rev. B \textbf{78}, 214402 (2008).
\bibitem{fe3o4_apl} P. Li., Y. Jiang, and H. L. Bai, Appl. Phys. Lett \textbf{96}, 092502 (2010).
\bibitem{fe3o4_jap} P. Li, C. Jin, E. Y. Jiang, and H. L. Bai, J. Appl. Phys \textbf{108}, 093921 (2010).
\bibitem{BKY04} Y. Bason, L. Klein, J.-B. Yau, X. Hong and C. H. Ahn, Appl. Phys. Lett. \textbf{84}, 2593 (2004).
\bibitem{BKY06} Y. Bason, L. Klein, J.-B. Yau, X. Hong, J. Hoffman and C. H. Ahn J. Appl. Phys \textbf{99}, 08R701 (2006).
\bibitem{BKW07} Y. Bason, L. Klein, H.Q. Wang, J. Hoffman, X. Hong, V.E. Henrich and C.H. Ahn, J. Appl. Phys \textbf{101}, 09J507 (2007).
\bibitem{Anisotropy_field1} J. A. Moyer, C. A. F. Vaz, E. Negusse, D. A. Arena, and V. E. Henrich, Phys. Rev. B \textbf{83}, 035121 (2011).
\bibitem{ZB00} M. Ziese and H. J. Blythe, J. Phys.: Condens. Matter \textbf{12}, 13 (2000).
\bibitem{Verwey_film} J.-B. Moussy, S. Gota, A. Bataille, M.-J. Guittet, M. Gautier-Soyer, F. Delille, B. Dieny, F. Ott, T. D. Doan, P. Warin, P. Bayle-Guillemaud, C. Gatel, and E. Snoeck. Phys. Rev. B, \textbf{70}, 174448 (2004).
\bibitem{Anisotropy_field2} L. Horng, G. Chern, M. C. Chen, P. C. Kang, and D. S. Lee, J. Magn Magn. Mater \textbf{270}, 389 (2004).
\bibitem{calcAPB} W. Eerenstein, T. T. M. Palstra, S. S. Saxena, and T. Hibma, Phys. Rev. Lett. \textbf{88}, 247204 (2002).
\bibitem{VAH09} C. A. F. Vaz, C. H. Ahn and V. E. Henrich in Epitaxial ferromagnetic films and spintronic applications, Research Signpost, (2009), p. 145.
\bibitem{new} D. Ihle and B. Lorenz, J. Phys. C: Solid State Phys. \textbf{19} (1986) 5239.
\bibitem{Sofin_MR_APB} R. G. S. Sofin, S. K. Arora, and I. V. Shvets, Phys. Rev. B \textbf{83}, 134436 (2011).
\bibitem{APB} D. T Margulies, F. T. Parker, M. L. Rudee, F. E. Spada, J. N. Chapman, P. R. Aitchison, and A. E. Berkowitz, Phys. Rev. Lett. \textbf{79}, 5162 (1997).
\bibitem{APDdistribution} A. M. Bataille, L. Ponson, S. Gota, L. Barbier, D. Bonamy, M. Gautier-Soyer, C. Gatel, and E. Snoeck, Phys. Rev. B \textbf{74}, 155438 (2006).
\bibitem{1Denergy} H. Zijlstra, IEEE Trans. Magn. \textbf{15}, 1246 (1979).
\bibitem{exchange} F. C. Voogt, T. T. M. Palstra, L. Niesen, O. C. Rogojanu, M. A. James, and T. Hibma, Phys. Rev. B \textbf{57}, R8107 (1998).
\bibitem{domain_size} W. Eerenstein, T. T. M. Palstra, T. Hibma, and S. Celotto, Phys. Rev. B \textbf{68}, 014428 (2003).



\end{thebibliography}
\end{document}